\documentstyle[emulateapj,apjfonts,psfig,graphicx]{article}

\def\grb{GRB\thinspace{980703}}

\begin{document}
 
\title{\large THE HOST GALAXY OF GRB\,980703 AT RADIO
WAVELENGTHS --- A NUCLEAR STARBURST IN A ULIRG}

\author{
E. Berger\altaffilmark{1},
S. R. Kulkarni\altaffilmark{1}, 
D. A. Frail\altaffilmark{2}
}

\altaffiltext{1}{Division of Physics, Mathematics, and Astronomy, 
California Institute of Technology 105-24, Pasadena, CA 91125}

\altaffiltext{2}{National Radio Astronomy Observatory, P.~O.~Box O,
  Socorro, NM 87801}

\begin{abstract}
\noindent We present radio observations of \grb{} at 1.43, 4.86, and 8.46 GHz
for the period of 350 to 1000 days after the burst.  These radio data
clearly indicate that there is a persistent source at the position of
\grb{} with a flux density of approximately 70 $\mu$Jy at 1.43 GHz,
and a spectral index, $\beta\approx 0.32$, where $F_\nu\propto
\nu^{-\beta}$.  We show that emission from the afterglow of \grb{} is
expected to be one to two orders of magnitude fainter, and therefore
cannot account for these observations.  We interpret this persistent emission as
coming from the host galaxy --- the first example of a $\gamma$-ray
burst (GRB) host detection at radio wavelengths.  We show that
emission from an AGN is unlikely, and find that it can
be explained as a result of a star-formation rate (SFR) of massive
stars ($M\!>\!5$ M$_\odot$) of $\approx 90$ M$_\odot$/yr, which gives
a total SFR (0.1M$_\odot<$M$<100$M$_\odot$) of $\approx 500$
M$_\odot$/yr. Using the correlation between the radio and far-IR (FIR)
luminosities of star-forming galaxies, we find that the host of \grb{}
is at the faint end of the class of Ultra Luminous Infrared Galaxies
(ULIRGs), with $L_{FIR}\sim few\times 10^{12}$ L$_\odot$. From the
radio measurements of the offset between the burst and the host, and
the size of the host, we conclude that \grb{} occurred near the center
of the galaxy in a region of maximum star formation. A comparison of
the properties of this galaxy with radio and optical surveys at a
similar redshift ($z\approx 1$) reveals that the host of \grb{} is an
average star-forming galaxy.  This result has significant implications
for the potential use of a GRB-selected galaxy sample for the study of
galaxies and the IGM at high redshifts, especially using radio
observations, which are insensitive to extinction by dust and provide
an unbiased estimate of the SFR through the well-known radio-FIR correlation.

\end{abstract}


\section{Introduction}\label{sec:intro}

Recent studies of the properties and host galaxies of $\gamma$-ray
bursts (GRBs) reveal some indirect evidence for the link between GRBs
and star formation.
Optical measurements of the offset distribution of GRBs from their 
host centers appears to be consistent with the distribution of
collapsars in an exponential disk, but inconsistent with the expected
offset distribution of delayed binary mergers (\cite{bkd01}).
GRB\thinspace990705 is an illustrative example of this
result since HST images revealed that the burst was situated in a
spiral arm, just north of an apparent star forming region (Holland et
al. 2000; \cite{bkd01}).  The absence of optical afterglows from the
so-called ``dark GRBs'' (Djorgovski et al. 2001\nocite{dfk+99}) points
to the association of GRBs with heavily obscured, and possibly
star-forming, regions.  In addition, Galama \& Wijers
(2000\nocite{gw00}) claim high column densities toward several GRBs from
X-ray observations of afterglows

Consequently, one of the pressing questions in the study of GRB host
galaxies is whether they are representative of star-forming galaxies
at a similar redshift.  If they are, then the dust-penetrating power
of GRBs and their broadband afterglow emission offer a number of
unique diagnostics of their host galaxies: the obscured star formation
fraction, the ISM within the disk, the local environment of the burst
itself, and global and line-of-sight extinction, to name a few.  

\grb{}, which has one of the brightest (apparent magnitude) hosts to
date ($R\approx 22.6$ mag; \cite{bfk+98}; \cite{vgo+99}) offers an 
excellent opportunity for detailed studies.  The afterglow optical and
near-IR (NIR) lightcurves exhibited pronounced flattening
about 6 days after the burst and this was attributed to an underlying
bright host (\cite{bfk+98}; \cite{czg+99}; \cite{vgo+99}).
Djorgovski et al. (1998\nocite{dkb+98b}) undertook spectroscopic
observations of the host and obtained a redshift of 0.966.  Using
three different estimators ([OII], H$\alpha$ and 2800\AA{} UV continuum) 
of the star-formation rate (SFR), Djorgovski et al. (1998) inferred 
extinction-corrected SFR of 10 to 30 M$_\odot$/yr. 

Here we report radio observations of \grb{} covering the period
350--1000 days after the burst at three frequencies: 1.43, 4.86, and
8.46 GHz.  This burst has the distinction of being followed up for 1000
days; the previous record-holder was GRB 970508 (445 days; \cite{fwk00}). 
The organization of the paper is as follows.  We summarize the radio 
observations and data reduction in \S\ref{sec:obs}.   In
\S\ref{sec:ag} we show that the late time radio observations require a
steady component over and above the decaying afterglow observations.
We argue that this component is unlikely to arise from an AGN but is
instead due to star-formation.  In \S\ref{sec:sfr}, we infer the SFR from the
radio observations and compare and contrast this estimate to those derived
from optical observations.  Thanks to the high angular resolution and
accurate astrometry of radio observations we are able to derive an
accurate offset between the burst and the centroid of the host, as well
as constrain the size of the radio emitting region (\S\ref{sec:offset}).

\section{Radio Observations}
\label{sec:obs}
Very Large Array (VLA\footnotemark\footnotetext{The VLA is operated by 
the National Radio Astronomy Observatory (NRAO), a facility
of the National Science Foundation operated under cooperative
agreement by Associated Universities, Inc.}) 
observations of \grb{} were initiated on 1998, July 4.40 UT
at 4.86 GHz.  All observations were obtained in the standard continuum
mode with $2\times 50$ MHz contiguous bands.  We used the extra-galactic
sources J2330+110, J0010+109 and J0022+061 for phase calibration
and 3C48 (J0137+331) and 3C147 (J0542+498) for flux calibration.
We used the Astronomical Image Processing System (AIPS) for data
reduction. The fluxes presented here are the values measured on
the images at the position of the source.

Late-time observations (time after the burst, $t\gtrsim 350$ days)
were co-added over a period of a few to thirty days in order to
increase the overall sensitivity of each detection.  This is
appropriate since the expected change in the flux density from the
afterglow over a few days, several hundred days after the burst, is
negligible relative to the associated errors in the measurements.  A
log of the late-time observations and the flux density measurements
are summarized in Table~\ref{tab:vla}, and the lightcurves are shown
in Figure~\ref{fig:radiolc}.  A summary of the early radio
data, as well as broadband modeling is given in Berger et al. 
(2001\nocite{b+01}).

\section{Evidence for Host Galaxy Emission in the Radio Regime}
\label{sec:ag}

\begin{figure*} 
\centerline{\hbox{\psfig{figure=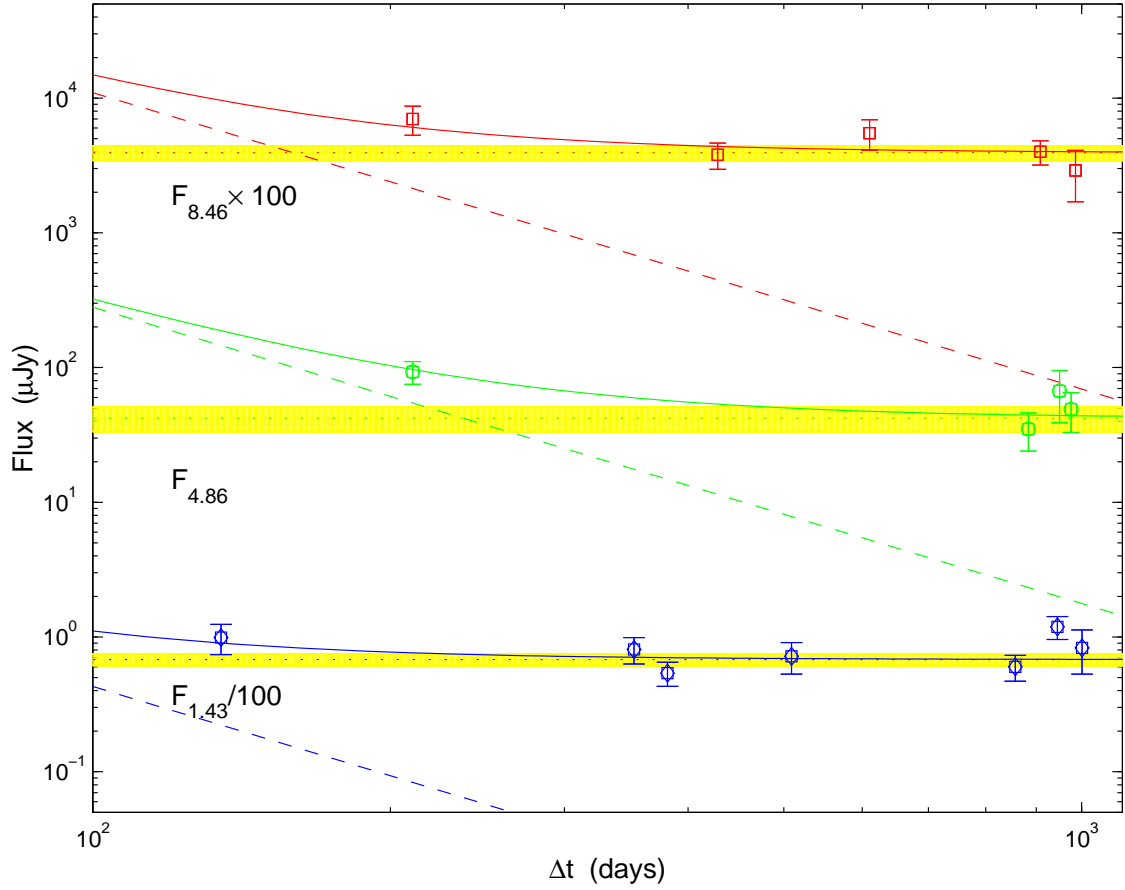,width=15cm}}}
\caption[]{Radio lightcurves at 1.43, 4.86, and 8.46 GHz.  The thin   
solid lines are the combined afterglow and host galaxy emission, the 
dotted lines indicate the afterglow emission, and the thick solid
lines are the weighted-average fluxes of the host galaxy, with the 
thickness indicating the uncertainty in the flux.  Only measurements
at $t\gtrsim 350$ days after the burst were used to calculate the host 
flux.  The fits are based on broadband fitting (Berger et al. 2001).
The data clearly indicate that there is a constant component in
the observed emission, interpreted as the host galaxy.  For details of 
the data reduction and analysis see section~\ref{sec:obs}.
\label{fig:radiolc}}
\end{figure*}

From Figure~\ref{fig:radiolc}, we see that the late-time ($t\gtrsim 350$
days) radio lightcurves do not exhibit the customary power-law decay 
expected of afterglows but instead show flattening.  From the early
broadband data we know that the afterglow spectrum peaked at
frequency, $\nu_m\sim 4\times 10^{12}$ Hz at $t=1.2$ days
(\cite{vgo+99}).  If the explosion was spherical then we expect
$\nu_m\propto t^{-3/2}$ (\cite{spn98}; \cite{cl00}).  Thus the radio
afterglow is expected to decay for $t>70$ days after the burst.  
If the ejecta were collimated (opening angle, $\theta_j$),
then we expect a more rapid decay, $\nu_m\propto t^{-2}$, once the
bulk Lorentz factor, $\Gamma$, of the flow falls below $\theta_j$,
$\Gamma(t)\lesssim \theta_j^{-1}$ (\cite{sph99}). In this case, we
expect the radio afterglow to start decaying at even earlier times,
and the flux will decay faster relative to a spherical explosion.  In 
either case, we expect the radio afterglow to decay by at least a
factor of three over the time span under consideration, $350<t<1000$ 
days.

We can clearly see from Figure~\ref{fig:radiolc} that this decay is
not taking place, and the flux instead remains constant over a period 
of approximately 650 days.  This behavior is similar to the flattening
observed in the optical/NIR lightcurves of several GRBs (including
\grb{}), when the emission from the afterglow decays below the level
of emission from the host galaxy.  

Furthermore, the afterglow spectrum is expected to be a power law,
$F_\nu\propto \nu^{-\beta}$, where $\beta=(p-1)/2$ and
$p$ is the power law index of the Lorentz factor distribution of the
shocked electrons, $N(\gamma)d\gamma \propto \gamma^{-p}d\gamma$ for 
$\gamma > \gamma_{\rm min}$ (\cite{spn98}).  From the observations of many
afterglows, we note that $p$ is in the range 2.2--2.6 and thus
we expect $\beta\sim 0.7$.  However, the observed spectral index
in the range 1.43--8.46 GHz is much lower, $\beta=0.32\pm 0.12$.  We
thus conclude that there exists a steady source of emission other than
the afterglow.

One possible explanation for this component is emission from an
active galactic nucleus (AGN).  It has been noted in surveys of
the Hubble Deep Field (HDF), its flanking fields, and the Small Selected
Area 13 (SSA13) that approximately 20\% of the radio sources are
AGN with spectral indices of about 0.3 (\cite{rfk+99};
\cite{r00}; \cite{bcr00}).  Windhorst et al. (1993) found a similar
result in their survey of two $7'\times 7'$ fields
at 8.44 GHz.  Thus, there is a modest probability
that the emission from the host of \grb{} is due to an AGN.  

We consider the AGN hypothesis unlikely based on the radio data and 
optical spectroscopy.  First, optical spectra of the source
obtained by Djorgovski et al. (1998) show no evidence for an
unobscured AGN: high-ionization lines such as Mg II $\lambda 2799$,
[NeV]$\lambda 3346$, and [NeV]$\lambda 3426$ are absent, and the
[OIII]$\lambda 4959$ to H$\beta$ ratio is approximately 0.4, much
lower than ${\rm [OIII]/H}\beta>1.3$ for AGN (\cite{rtt97}).  
Another way to discriminate between AGN and star-forming
galaxies is to correlate the [OII] equivalent width (EW) with
continuum color (\cite{dg82}).  Kennicutt (1992) showed that AGN have
redder colors for similar [OII] EW, relative to normal galaxies.  
Using the spectrum presented in Djorgovski et al. (1998) we evaluate
the color index, $(41-50)\equiv 2.5{\rm
log}[f_\nu(5000{\rm \AA})/f_\nu(4100{\rm \AA})]$ (\cite{k92}), and find it to be
$0\pm 0.1$; an AGN with the same [OII] EW would have a
value $\gtrsim 0.3$ (\cite{k92}).  Finally, Rola, Terlevich \& Terlevich
(1997\nocite{rtt97}) found that for a sample of emission-line galaxies 
at $z\sim 0.8$, the color index between the continuum
underlying the H$\beta$ and [OII]$\lambda 3727$ lines is $\ge 0.4$ for 
all AGN in their sample.  Using the spectrum of Djorgovski
et al. (1998) we find that this color index is approxmiately zero.

\begin{figure*} 
\centerline{\hbox{\psfig{figure=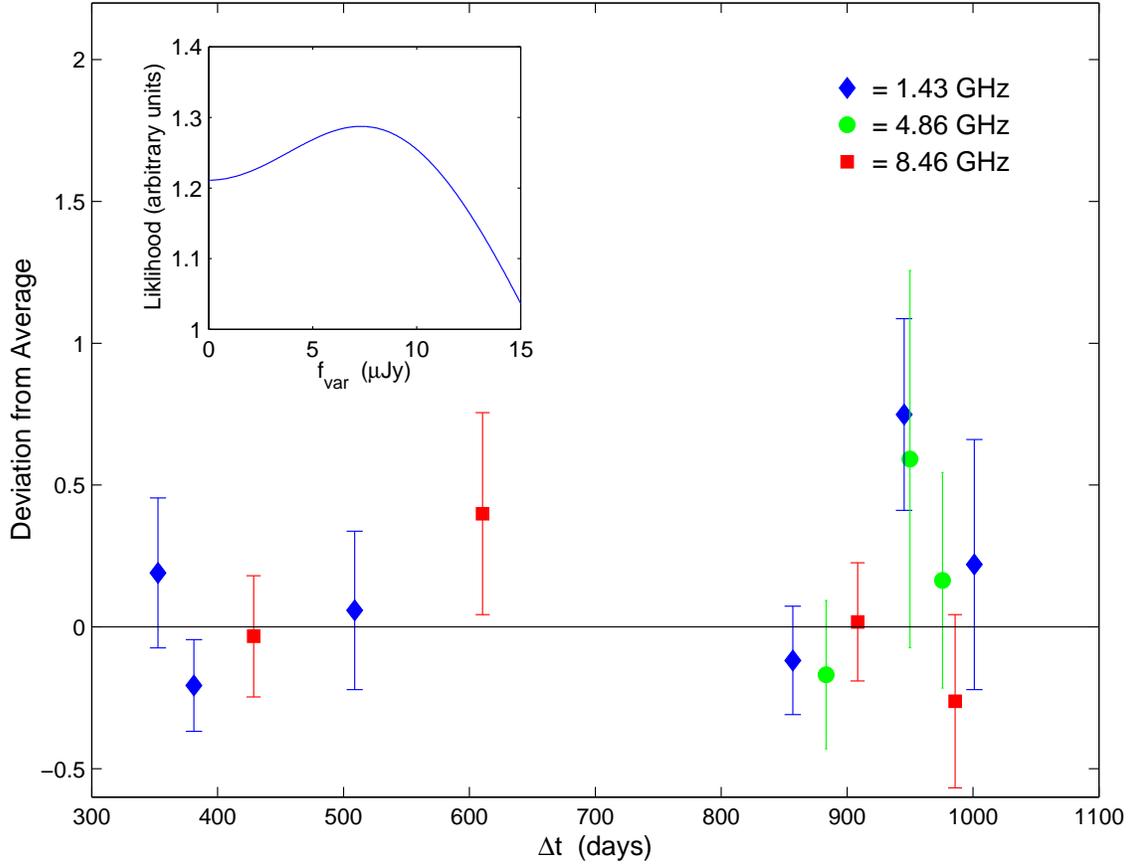,width=15cm}}}
\caption[]{Fluctuations of individual measurements around the weighted
average presented by the wide strips in figure~\ref{fig:radiolc}.  We
note that there are no fluctuations above 1.1$\sigma$ at 8.46 GHz,
0.8$\sigma$ at 4.86 GHz, and 1.7$\sigma$ at 1.43 GHz, indicating that 
the flux in each band is consistent with a constant.  In
fact, if we assume that the source has some variability (over that due
to measurement errors), then the variable flux is less than
$\pm 7$ $\mu$Jy, with a 40\% probability that there are no
fluctuations at all (see insert).  This conclusion supports the
hypothesis that the radio emission is not due to an AGN.
\label{fig:fluc}}
\end{figure*}

A second, but less persuasive argument against an AGN origin is the
apparent absence of significant radio variability over the 650 day
monitoring period (see Figure~\ref{fig:fluc}).  The radio cores of
most, but not all, low-luminosity AGN show variability exceeding the
observed levels (\cite{flb+00}). 

We thus conclude that the radio emission seen from the host of \grb{}
is unlikely to be due to AGN activity.  However, star-forming galaxies
exhibit radio emission arising from their supernova remnants (SNRs) and
HII regions. In the next section we show how the observations, the
radio spectral index, and the optical spectrum, are consistent with
the hypothesis that the radio emission is related to star formation.

\section{The Star-Formation Rate in the Host Galaxy of \grb{}}
\label{sec:sfr}

Star formation is traced by optical, far-IR, sub-mm, and radio
emission.  In the following we will use the radio data to estimate the
SFR in the host galaxy of \grb{}, and then compare the results with
the SFR derived from optical indicators, and with radio surveys at a
similar redshift range in order to place the host of \grb{} in a
larger context.

\subsection{Star Formation Rate from the Radio Observations}
\label{sec:sfrr}

Using all the measurements in Table~\ref{tab:vla}, we find
the following weighted-average flux densities for the host galaxy of 
\grb{}: $F_{\nu,8.46}=39.3\pm 4.9$ $\mu$Jy,
$F_{\nu,4.86}=42.1\pm 8.6$ $\mu$Jy, and $F_{\nu,1.43}=68.0\pm 6.6$
$\mu$Jy.  From the redshift of \grb{}, $z=0.966$ (\cite{dkb+98b}), and 
the cosmological parameters $\Omega_0=0.3$, $\Lambda_0=0.7$ and $H_0=65$
km/sec/Mpc, we find that the luminosity distance to the burst is
$d_L=d_A(1+z)^2\approx 2.1\times 10^{28}$ cm, and the observed
luminosity at each frequency is given by $L_\nu=4\pi d_L^2F_\nu$.
The emitted luminosity is given by
$L_{em,\nu'}=L_{obs,\nu}(\nu/\nu')^\beta(1+z)^\beta$, where $\beta$ is
the spectral index of the radio emission, $\nu'$ is the rest
frequency, and $\nu$ is the observing frequency.  From the flux
densities we find that the radio spectral index is
$\beta=0.32\pm 0.12$, and thus the emitted luminosity in each
frequency is approximately 25\% higher than the observed luminosity 
at the same frequency.  
 
Condon (1992) showed that the total luminosity is a combination of
synchrotron and thermal emission components, both directly related to 
the formation rate of massive stars via a simple relationship.
Moreover, since the lifetime of massive stars is of the order of
$10^7$ years, and the lifetime of the synchrotron emitting electrons
is of the order of $10^8$ years (\cite{c92}), the radio emission is an
excellent probe of the instantaneous SFR.  Using the emitted
luminosity at $\nu'=1.43$ GHz, $L_{em}(1.43)=(4.7\pm 0.6)\times
10^{30}$ erg sec$^{-1}$, we find that the SFR of massive stars in the
host of \grb{} is
\begin{equation}
{\rm SFR}({\rm M}\!>\!5{\rm M}_\odot)\approx \frac{L_{em}(1.43)}{5.3\times
10^{28}{\nu'}_{\rm GHz}^{-\beta}+5.5\times 10^{27}{\nu'}_{\rm
GHz}^{-0.1}}\approx 90 {\rm M}_\odot/{\rm yr}.
\label{eqn:massivesfr}   
\end{equation}

Since both the thermal and non-thermal components are proportional only
to the formation rate of high-mass stars,
equation~\ref{eqn:massivesfr} has to be modified by a factor which
accounts for the contribution from stars in the mass range 0.1--5
M$_\odot$.  For a Salpeter IMF this factor evaluates to 5.5.  We use
the Salpeter IMF since it is already implicitly used in
equation~\ref{eqn:massivesfr} for the mass range 5--100 M$_\odot$.    
Thus, within this framework the total SFR is $\approx 500$ M$_\odot$/yr.

We now compare the SFR and other characteristics of the host of \grb{} 
with those of other galaxies at a similar redshift.
A radio survey of the Hubble Deep Field (HDF) and its flanking fields
showed that the mean spectral index of the 8.46 GHz selected sample is
$\langle\beta_{8.46}\rangle=0.35\pm 0.07$ (\cite{r00}).  In a survey of two 
$7'\times 7'$ fields with the VLA at 8.44 GHz Windhorst et al. (1993)
found for sources with a flux density $\lesssim 100$ $\mu$Jy
(hereafter, $\mu$Jy sources) a median spectral index, $\beta_{\rm
med}\approx 0.35\pm 0.15$, and Fomalont et al. (1991) found 
$\beta_{\rm med}\approx 0.38$ for $\mu$Jy sources selected at 4.9 GHz.
Thus, the host galaxy of \grb{} appears to be a normal $\mu$Jy source
compared to sources selected at 4.9 or 8.5 GHz.
In addition, it has been noted (\cite{r00}) that the spectral index 
of radio sources selected at frequencies larger than 5 GHz flattens
from a value of approximately 0.7 for the mJy ($F_\nu\gtrsim 1$ mJy)
population to 0.3 for $\mu$Jy sources.

The reason for the flattening of the spectral index is a varying ratio
of thermal bremsstrahlung to synchrotron emission.  Supernova remnant
shock acceleration of electrons results in synchrotron emission, with
a characteristic spectral index of $\approx 0.8$ (\cite{c92}).  On the
other hand, thermal bremsstrahlung emission from HII regions, excited
by star formation, has a much flatter spectral index, $\beta\approx 0.1$.
Thus, as the direct contribution from massive stars increases the
spectra are expected to flatten from a value of 0.8 to 0.1.  This is
exactly the effect that is observed in the aforementioned surveys.  

Within the HDF and SSA13 Richards et al. (1999\nocite{rfk+99})
identified radio sources with fluxes in the range 10--100
$\mu$Jy with bright disk galaxies with $I\approx 22$ mag.  The $I-K$
color for these galaxies is approximately 2.5 mag.  Bloom et
al. (1998) find $I\approx 21.9$ mag and $I-K\approx 2.1$ mag for the
host of \grb{}.  Thus, we see from both the radio spectrum of the
source, and the optical $I$ mag and $I-K$ color that the host galaxy
of \grb{} has the characteristics of a typical star-forming radio
galaxy selected at 8.5 GHz.  

\begin{figure*} 
\centerline{\hbox{\psfig{figure=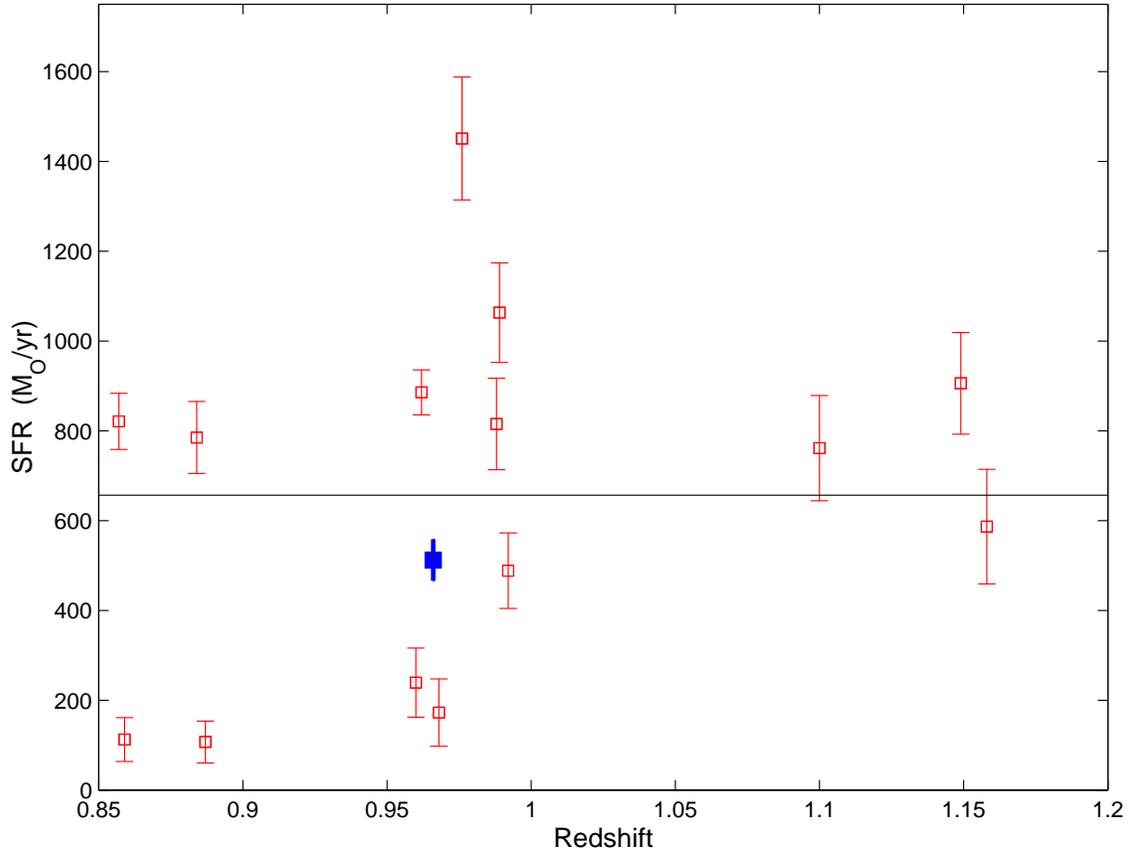,width=15cm}}}
\caption[]{SFR for radio galaxies in the HDF, SSA13 and V15 (Haarsma et
al. 2000) in the redshift range 0.85--1.15.  The dashed line is an
average over this redshift range.  The host galaxy of \grb{} is marked
by a solid square.  The errorbars on the survey sources indicate a 10
$\mu$Jy measurement error.  The host of \grb{} appears to be an
average star-forming galaxy relative to the sample. 
\label{fig:hdf}}
\end{figure*}

Figure~\ref{fig:hdf} shows the total SFR in the host galaxy of \grb{}
as compared to sources in the HDF, its flanking fields, SSA13, and V15
(\cite{hpw+00}) in the redshift range $0.85-1.15$.  These fields
have been observed to $\mu$Jy sensitivities at cm wavelengths, and
the detected radio sources have been identified with optical sources
for which the redshift was determined.  We use the flux and spectral
index measurements along with equation~\ref{eqn:massivesfr} and the
correction factor to calculate the total SFR.  It is 
clearly
seen from the figure that the host of \grb{} is an average galaxy at
$z\approx 1$ based on star formation.  This conclusion meshes well
with the comparison of the radio spectral index, optical $I$ mag, and
optical $I-K$ color of the host of \grb{} to the same sample.

\subsection{Star Formation Rate from Optical and Sub-mm Data}
\label{sec:sfrm}

Djorgovski et al. (1998) used H$\alpha$ and the 2800\AA{} UV continuum 
to calculate a SFR of approximately 10 M$_\odot$/yr in the host of
\grb{}, after correcting for rest-frame extinction, $A_V\approx 0.3$
mag.  Sokolov et al. (2001) found a similar intrinsic extinction, 
$A_V\approx 0.3-0.65$, and based on template spectral energy
distributions found that the best model for the broadband optical
spectrum is given by exponentially decreasing star formation with an 
extinction-corrected SFR of 20 M$_\odot$/yr.

Clearly, the SFR derived from optical indicators is much
lower than the value from radio measurements, even after correcting
for extinction.  This result is part of a general trend that has been
observed in galaxies with ${\rm SFR}\gtrsim 0.1$ M$_\odot$/yr
(\cite{hch+01}).  Hopkins et al. (2001) propose dust reddening
dependent on SFR as the solution to this problem, and we therefore
expect a much better result if we use their prescription.  Extending 
their correlation to ${\rm SFR}_{1.43}\approx 500$ M$_\odot$/yr, we 
find that the predicted observed SFR from H$\alpha$ is approximately 
70 M$_\odot$/yr.  This value is still much higher than the measured
SFR.  In fact, in the Hopkins et al. (2001) 
sample the optically-derived SFR rarely exceeds 10 M$_\odot$/yr,
while the radio-derived values go up to several hundred M$_\odot$/yr,
indicating that the optical emission does not trace the entire
star-forming volume.  Thus, the SFR values for this particular galaxy 
are not unexpected. 

The sub-mm (e.g. 350 GHz) emission from galaxies serves as another
estimator of SFR,
and it is related to the radio emission at 1.43 GHz via a
redshift-dependent spectral index, $\beta_{1.4}^{350}$ (\cite{cy99},
2000\nocite{cy00}; \cite{dce00}).  Using the calibration of Dunne,
Clements \& Eales (2000) we find a value of $\beta_{1.4}^{350}\approx
0.6\pm 0.1$ at $z\approx 1$, which gives $F_{\nu}(350)\approx
1.9^{+1.4}_{-0.8}$ mJy.  Observations with the Sub-millimeter Common
User Bolometer Array (SCUBA) camera on the James Clark Maxwell
Telescope (JCMT) 12.4 days after the burst provided a $2\sigma$ upper
limit of 3.2 mJy on the combined emission from the afterglow and host
at 350 GHz (\cite{stv99}), consistent with the predictions from the 
radio--sub-mm relation.  
 
To conclude, we use the derived SFR to calculate the expected far-IR
(FIR) emission from the host of \grb{}.  The luminosity of the FIR
radiation can be derived from the empirical relation suggested by
Helou, Soifer \& Rowan-Robinson (1985\nocite{hsr85}),
\begin{equation}
q=-12.6+{\rm log}(F_{FIR}/F_{1.4})\approx 2.3
\label{eqn:firsfr}
\end{equation}
which evaluates to $L_{FIR}\approx 10^{12}$ L$_\odot$ for the host of
\grb{}; here $F_{FIR}$
is the total flux in the range 40--120 $\mu$m in units of erg sec$^{-1}$
cm$^{-2}$, and $F_{1.4}$ is the flux density at 1.4 GHz in units of
erg sec$^{-1}$ cm$^{-2}$ Hz$^{-1}$.  
Hopkins et al. (2001) provide the calibration $L_{FIR}=5.81\times
10^9\times {\rm SFR}$ L$_\odot$, which gives $L_{FIR}\approx 3\times
10^{12}$ L$_\odot$, in good agreement with the previous value.  
These values of the FIR luminosity place the host galaxy of \grb{} in
the category of ULIRG (\cite{sm96}).

\section{Offset Measurements and Source size}
\label{sec:offset}

\begin{figure*} 
\centerline{\hbox{\psfig{figure=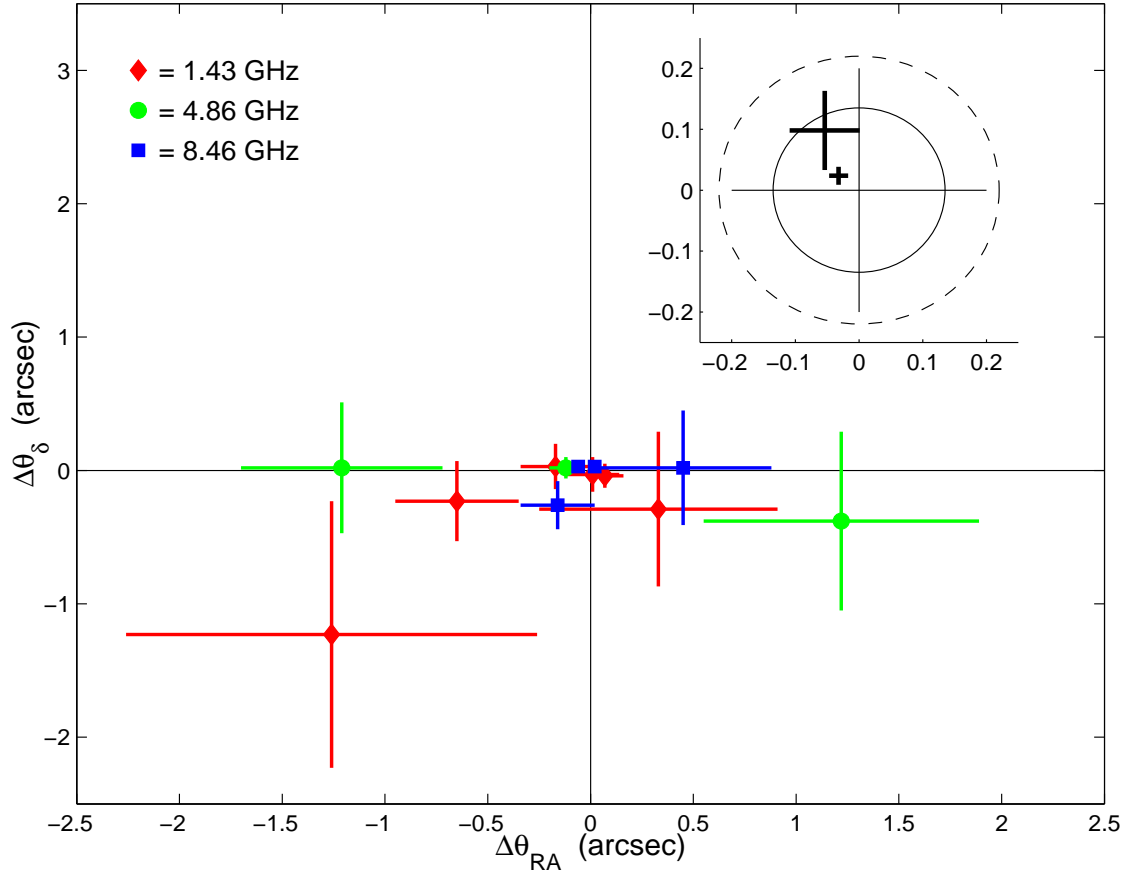,width=15cm}}} 
\caption[]{Offset measurements for all epochs in which the host galaxy 
emission dominates.  The plot shows the offset in RA and $\delta$ of
the VLBA position of the burst (see section~\ref{sec:offset}) relative to 
the host center, ($\Delta\theta_{\rm RA},\Delta\theta_\delta$)=(0,0).
The most accurate measurements are at 8.46 GHz in the VLA A
configuration.  In this mode we achieved an rms positional error of
0.02 arcsec.  The insert shows the weighted average offset in both RA
and $\delta$ (small cross).  The larger cross is the offset
measurement from Bloom, Kulkarni, and Djorgovski (2001).  The solid
circle designates the projected maximum source size from the radio
observations in the A configuration at 8.46 GHz, and the dashed circle 
is the optical size from Holland et al. (2001).  Clearly the formation
of massive stars is concentrated in the central region of the host.  The
small offset of the burst from the host center indicates that \grb{}
occurred in the region of maximum star formation, which points to a
link between GRBs and massive stars.
\label{fig:offset}}
\end{figure*}

Figure~\ref{fig:offset} shows the projected angular offset between the
host galaxy and afterglow of \grb, for each individual detection, and
the combined value from all observing runs (insert in
Figure~\ref{fig:offset}).  Positions are determined from Gaussian
fits, and the host-GRB offset is calculated with respect to a Very
Long Baseline Array (VLBA) position that was measured to 0.0007 arcsec
accuracy in each coordinate on 1998 August 2 at 8.42 GHz (Berger et
al. 2001\nocite{b+01}).

We find an average offset from all measurements of $-0.032\pm 0.015$
arcsec in RA and $0.024\pm 0.015$ arcsec in declination.  The
uncertainty in the position of the source is given by
$\delta\theta_{\rm offset}\approx (\theta_{syn beam}/2)/(S/N)$,
where $\theta_{syn beam}\approx \lambda/B_{max}$ is the half-power
synthesized beam-width, $\lambda$ is the observing wavelength, $B_{max}$ is
the length of the maximum baseline, and $S/N$ is the signal-to-noise
ratio of the flux measurement.

The optical measurements of Bloom, Kulkarni \& Djorgovski
(2000) for the host of \grb{}, give an angular offset of $-0.054\pm
0.055$ in RA and $0.098\pm 0.065$ in declination (see insert in
Figure~\ref{fig:offset}).  They
conclude that \grb{} was not significantly offset from the center of
its host galaxy, in agreement with the more
accurate offset measurements in the radio.
 
In addition to accurate measurements of the offset, the radio observations
allow us to place meaningful limits on the size of the radio-emitting
region (i.e. the size of the star-forming region).  We find that in
our highest resolution images the source is unresolved, and therefore,
based on the synthesized beam size we can derive an upper limit on the
physical size of the source.  For our adopted cosmological parameters 
(section~\ref{sec:sfr}) we find that the angular diameter distance to
the source is $d_A\approx 5.4\times 10^{27}$ cm.  The full synthesized
beam-width at 8.46 GHz is $\theta_{HPBW}\approx 0.27$ arcsec, which
gives an upper limit of $D_{rad}=d_A\theta_{HPBW}<2.3$ kpc on the 
diameter of the source.  

Holland et al. (2001\nocite{hfh+01}) used a $R^{1/n}$ profile to fit
the optical emission from the host and found that the best fit gives a 
half-light radius of 0.13 arcsec, which corresponds to an exponential
disk with a scale-diameter of 0.44 arcsec.  Thus, the physical size of 
the galaxy is $D_{opt}\approx 3.7$ kpc, 60\% larger than the upper
limit from our radio measurements.  However, Holland et al. (2001)
claim that the center of the galaxy is 0.2 mag bluer than the outer
regions of the host.   If so, star formation must be mainly taking
place within the inner parts of the galaxy.  Since the radio emission
directly traces current star formation, we expect the radio emission
to be more centrally concentrated than the optical emission.  Thus, as
expected, the radio size of the galaxy is smaller than the optical size.  

Most likely the GRB is located within the nuclear starburst given the 
small offset of the GRB from the centroid of the galaxy. If so it raises
the question of why the afterglow was not completely
extinguished by dust.  In fact, in order to reconcile the
optical and radio derived SFRs we require a rest-frame extinction of
$A_V\sim 4.5$ mag.  Observations of the afterglow, which provide an
estimate of extinction along the line-of-sight to the burst, give values
of 1--2 mags from optical observations, and somewhat higher values from
X-ray observations (\cite{bfk+98}; \cite{czg+99}; \cite{vgo+99}).
Thus, the extinction in the nuclear star-forming region is higher than
the average over the whole galaxy, and the correction to the observed
optical SFR is almost sufficient to reconcile it with the value of 500
M$_\odot$/yr derived from the radio.

The relatively small source size also agrees well with the
classification of the host of \grb{} as a ULIRG exhibiting a
starburst.  Kennicutt (1998, and references therein) showed that star
formation with a rate $\gtrsim 20$ M$_\odot$/yr invariably takes place
in circumnuclear regions of size 0.2--2 kpc, in the form of nuclear
starburst.  As a result, we expect that ULIRGs will have such size
scales when traced by star formation, and the source size we measured
for the host of \grb{} indicates that it is probably undergoing a
nuclear starburst.  

Finally, from the source size and offset measurement we conclude that 
\grb{} took place in the region of maximum star formation, providing
further indirect evidence linking GRBs to massive stars

\section{Conclusions}
\label{sec:conc}

Late-time observations of \grb{} reveal a steady component, with a
flux density $F_{\nu,1.43}=68.0\pm 6.6$ $\mu$Jy and a spectral index
$\beta=0.32\pm 0.12$.  The spectral and temporal characteristics of
this emission indicate that it does not arise from the afterglow
itself, but rather it is the result of star formation in the host
galaxy, with ${\rm SFR}\approx 500$ M$_\odot$/yr.  This leads to the
interpretation that this host galaxy is a ULIRG undergoing a 
starburst.  In addition, the star formation is concentrated within the
inner two kpc of the host, and the progenitor of \grb{} was positioned
within this region of star formation.  This conclusion lends
additional support for the collapsar model.  

If GRBs really come from massive stars, then they can be used to trace 
the star formation history of the universe (e.g. \cite{bn00}).
In addition, GRBs and their afterglows are potentially detectable 
out to very high redshifts (\cite{lr00}).  These propositions, taken
together with the dust-penetrating power of their $\gamma$-ray
emission, make GRBs a unique tool for the study of galaxies
and the IGM over a wide redshift range.  In particular, radio and 
sub-mm/FIR observations of a GRB-selected galaxy sample will be
extremely useful for the study of the obscured star formation
fraction, and the properties of starbursts at high redshifts.
Moreover, a comparison of the global star formation history as derived
from these long-wavelength host studies, with the redshift
distribution of GRBs, will provide valuable insight as to how
well GRBs trace the formation rate of massive stars; we expect that
if GRBs trace only a particular channel of star formation, the two
distributions will not agree.  

Therefore, it is imperative to study the hosts of GRBs in the radio
and sub-mm/FIR.  Future observatories such as the Space Infrared
Telescope Facility (SIRTF; to be launched in July 2002), the Expanded
VLA (EVLA), and the Square-Kilometer Array (SKA) will allow detailed
studies of these hosts.  In the FIR, SIRTF will have the ability
to detect sources down to a few mJy, allowing the detection of
galaxies with SFR comparable to that in the host of \grb{} out to
$z\sim 10$; alternatively, we will be able to detect hosts with SFR as
low as a few M$_\odot$/yr at $z\sim 1$.  The EVLA and SKA will greatly 
improve the detectability of host galaxies in the radio, and will also
allow much higher angular resolution studies of compact star-forming
regions.  With a factor ten increase in resolution and a factor five
increase in sensitivity over the current VLA, we will be able to probe
scales of approximately 5 mas with the EVLA; for a galaxy at $z\approx
1$ this translates to a physical scale of 150 pc.  In addition,
EVLA will detect galaxies with a total SFR as low as 50 M$_\odot$/yr
at $z\sim 1$.  The SKA, with a similar resolution, but a much larger
collecting area, will extend this capability to even lower SFR and
smaller star-forming regions.   

Thus, as more host galaxies are detected and studied in detail in the
radio and sub-mm/FIR, we will be able to address a large number of
issues pertaining not only to the bursts themselves, but also to the
characteristics of galaxies at high redshifts.

\acknowledgements We acknowledge support by NSF and NASA grants.


\begin{thebibliography}{}

\bibitem[{Barger}, {Cowie} \& {Richards} 2000]{bcr00}
{Barger}, A.~J., {Cowie}, L.~L., \& {Richards}, E.~A. 2000, \aj, 119, 2092.

\bibitem[{Berger} {et al.}  2001]{b+01}
{Berger}, E., {et al.} 2001, in preparation.

\bibitem[{Blain} \& {Natarajan} 2000]{bn00}
{Blain}, A.~W. \& {Natarajan}, P. 2000, MNRAS, 312, L35.

\bibitem[{Bloom} {et al.}  1998]{bfk+98}
{Bloom}, J.~S., {et al.}  1998, ApJ, 508, L21.

\bibitem[{Bloom}, {Kulkarni} \& {Djorgovski} 2001]{bkd01}
{Bloom}, J.~S., {Kulkarni}, S.~R., \& {Djorgovski}, S.~G. 2001, submitted to AJ.
  astro-ph/0010176.

\bibitem[{Carilli} \& {Yun} 1999]{cy99}
{Carilli}, C.~L. \& {Yun}, M.~S. 1999, \apjl, 513, L13.

\bibitem[{Carilli} \& {Yun} 2000]{cy00}
{Carilli}, C.~L. \& {Yun}, M.~S. 2000, \apj, 530, 618.

\bibitem[{Castro-Tirado} {et al.}  1999]{czg+99}
{Castro-Tirado}, A.~J., {et al.}  1999, ApJ, 511, L85.

\bibitem[{Chevalier} \& {Li} 2000]{cl00}
{Chevalier}, R.~A. \& {Li}, Z. 2000, ApJ, 536, 195.

\bibitem[{Condon} 1992]{c92}
{Condon}, J.~J. 1992, \araa, 30, 575.

\bibitem[{Djorgovski} {et al.}  2001]{dfk+99}
{Djorgovski}, S.~G., {et al.}  2001, in preparation.

\bibitem[{Djorgovski} {et al.}  1998]{dkb+98b}
{Djorgovski}, S.~G., {et al.}  1998, ApJ, 508, L17.

\bibitem[{Dressler} \& {Gunn} 1982]{dg82}
{Dressler}, A. \& {Gunn}, J.~E. 1982, ApJ, 263, 533.

\bibitem[{Dunne}, {Clements} \& {Eales} 2000]{dce00}
{Dunne}, L., {Clements}, D.~L., \& {Eales}, S.~A. 2000, \mnras, 319, 813.

\bibitem[Falcke {et al.}  2000]{flb+00}
{Falcke}, H., {et al.} 2000, {to appear in "Probing the Physics of
Active Galactic Nuclei by Multiwavelength Monitoring",
eds. B.M. Peterson, R.S. Polidan, \& R.W. Pogge, ASP Conf. Ser.
astro-ph/0009457}.

\bibitem[{Fomalont} {et al.}  1991]{fwk+91}
{Fomalont}, E.~B., {et al.} 1991, AJ, 102, 1258.

\bibitem[{Frail}, {Waxman} \& {Kulkarni} 2000]{fwk00}
{Frail}, D.~A., {Waxman}, E., \& {Kulkarni}, S.~R. 2000, ApJ, 537, 191.

\bibitem[{Galama} \& {Wijers} 2000]{gw00}
{Galama}, T.~J. \& {Wijers}, R. A. M.~J. 2000, ApJ in press; astro-ph/0009367.

\bibitem[{Haarsma} {et al.}  2000]{hpw+00}
{Haarsma}, D.~B., {et al.}  2000, \apj, 544, 641.

\bibitem[{Helou}, {Soifer} \& {Rowan-Robinson} 1985]{hsr85}
{Helou}, G., {Soifer}, B.~T., \& {Rowan-Robinson}, M. 1985, \apjl, 298, L7.

\bibitem[Holland {et al.}  2001]{hfh+01}
{Holland}, S. {et al.}  2001, submitted to A\&A. astro-ph/0103058.

\bibitem[{Hopkins} {et al.}  2001]{hch+01}
{Hopkins}, A.~M., {et al.}  2001, accepted to AJ. astro-ph/0103253.

\bibitem[{Kennicutt} 1992]{k92}
{Kennicutt}, R.~C. 1992, \apj, 388, 310.

\bibitem[Kennicutt(1998)]{k98}
Kennicutt, R.~C. 1998, \araa, 36, 131.

\bibitem[{Lamb} \& {Reichart} 2000]{lr00}
{Lamb}, D.~Q. \& {Reichart}, D.~E. 2000, \apj, 536, 1.

\bibitem[{Richards} 2000]{r00}
{Richards}, E.~A. 2000, \pasp, 112, 1001.

\bibitem[{Richards} {et al.}  1999]{rfk+99}
{Richards}, E.~A., {et al.}  1999, \apjl, 526, L73.

\bibitem[{Rola}, {Terlevich} \& {Terlevich} 1997]{rtt97}
{Rola}, C.~S., {Terlevich}, E., \& {Terlevich}, R.~J. 1997, \mnras, 289, 419.

\bibitem[{Sanders} \& {Mirabel} 1996]{sm96}
{Sanders}, D.~B. \& {Mirabel}, I.~F. 1996, \araa, 34, 749+.

\bibitem[{Sari}, {Piran} \& {Halpern} 1999]{sph99}
{Sari}, R., {Piran}, T., \& {Halpern}, J.~P. 1999, ApJ, 519, L17.

\bibitem[{Sari}, {Piran} \& {Narayan} 1998]{spn98}
{Sari}, R., {Piran}, T., \& {Narayan}, R. 1998, ApJ, 497, L17.

\bibitem[{Smith} {et al.}  1999]{stv99}
{Smith}, I.~A. {et al.}  1999, \aap, 347, 92.

\bibitem[{Sokolov} {et al.} 2001]{sfc+01}
{Sokolov}, V.~V. {et al.}  2001, submitted to A\&A; astro-ph/0104102.

\bibitem[{Vreeswijk} {et al.}  1999]{vgo+99}
{Vreeswijk}, P.~M. {et al.}  1999, ApJ, 523, 171.

\bibitem[{Windhorst} {et al.} 1993]{wfp+93}
{Windhorst}, R.~A. {et al.}  1993, ApJ, 405, 498.


\end{thebibliography}

\begin{deluxetable}{llcccc}
\tabcolsep0in\footnotesize
\tablewidth{\hsize}
\tablecaption{Radio Observations of \grb\ \label{tab:vla}}
\tablehead {
\colhead {Epoch}      &
\colhead {$\Delta t$} &
\colhead {Array Configuration} &
\colhead {$t^{\rm on-source}$} &
\colhead {$\nu_0$} &
\colhead {S$\pm\sigma$} \\
\colhead {(UT)}      &
\colhead {(days)} &
\colhead {} &
\colhead {(hrs)} &
\colhead {(GHz)} &
\colhead {($\mu$Jy)}
}
\startdata
1999 June $\phantom{0}$15.36---26.29  & 352.65 & A & 6.5  & 1.43 & 81$\pm$18 \nl
1999 July $\phantom{0}$10.53---28.28  & 381.22 & A   & 13.4 & 1.43 & 57$\pm$10 \nl
1999 August $\phantom{0}$19.40---September $\phantom{0}$21.24 & 428.64 & A & 11.3 & 8.46 & 37$\pm$8  \nl
1999 November $\phantom{0}$24.06 & 508.88  & B & 1.7 & 1.43 & 57$\pm$20  \nl
2000 March $\phantom{0}$5.70  & 610.52  & BnC & 2.8 & 8.46 & 57$\pm$14  \nl
2000 October $\phantom{0}$7.30---November $\phantom{0}$19.08 & 847.01 & A & 6.5 & 1.43 & 76$\pm$11 \nl
2000 December $\phantom{0}$2.19---4.97    & 882.60 & A & 4.9 & 4.86 & 35$\pm$11 \nl
2000 December $\phantom{0}$21.15---2001 January $\phantom{0}$4.98 & 908.38 & A & 7.0 & 8.46 & 41$\pm$8  \nl
2001 February $\phantom{0}$2.00---4.93 & 945.29 & AnB  & 1.9 & 1.43 & 113$\pm$21  \nl
2001 February $\phantom{0}$8.08 & 949.90 & AnB & 1.2 & 4.86 & 57$\pm$28  \nl
2001 March $\phantom{0}$2.98---9.00 & 975.81 & B & 4.5 & 4.86 & 43$\pm$15  \nl
2001 March $\phantom{0}$22.96---April $\phantom{0}$8.56 & 1001.08 & B & 2.6 & 1.43 & 83$\pm$30 \nl  
\enddata
\tablecomments{The columns are (left to right), (1) UT date of the
start of each observation or range of dates for observations which
were added over several days, (2) time elapsed since the $\gamma$-ray
burst, (3) array configuration, (4) total on-source observing time,
(5) observing frequency, and (6) peak flux density at the best fit
position of the radio transient, with the error given as the root mean
square noise on the image.}
\end{deluxetable}

\clearpage 

\clearpage

\clearpage

\clearpage 

\end{document}